\documentclass[preprintnumbers,amsmath,amssymb,groupedaddress,superscriptaddress,prr]{revtex4-1}

\usepackage[utf8]{inputenc}
\usepackage{graphicx}
\usepackage{dcolumn}
\usepackage[colorlinks=true,citecolor=blue,urlcolor=blue]{hyperref}
\usepackage[dvipsnames]{xcolor}
\usepackage{subfigure}

\usepackage{amsmath,amssymb,cases} 
\usepackage{bm}

\begin{document}
\title{A Dual-Species Bose-Einstein Condensate with Attractive Interspecies Interactions}
\author{A. Burchianti}\email{burchianti@lens.unifi.it}
\affiliation{Istituto Nazionale di Ottica, CNR-INO, 50019 Sesto Fiorentino, Italy}
\affiliation{\mbox{LENS and Dipartimento di Fisica e Astronomia, Universit\`{a} di Firenze, 50019 Sesto Fiorentino, Italy}}
\author{C. D'Errico}\altaffiliation{Present address: Istituto per la Protezione Sostenibile delle Piante, CNR‐IPSP, 10135 Torino, Italy}
\affiliation{Istituto Nazionale di Ottica, CNR-INO, 50019 Sesto Fiorentino, Italy}
\affiliation{\mbox{LENS and Dipartimento di Fisica e Astronomia, Universit\`{a} di Firenze, 50019 Sesto Fiorentino, Italy}}
\altaffiliation{Present address: Istituto per la Protezione Sostenibile delle Piante, CNR‐IPSP, 10135 Torino, Italy}
\author{M. Prevedelli}
\affiliation{Dipartimento di Fisica e Astronomia, Universit\`{a} di Bologna, 40127 Bologna, Italy}
\author{L. Salasnich}
\affiliation{Istituto Nazionale di Ottica, CNR-INO, 50019 Sesto Fiorentino, Italy}
\affiliation{\mbox{Dipartimento di Fisica e Astronomia 'Galileo Galilei' and CNISM, Universit\`{a} di Padova, 35131 Padova, Italy}}
\author{F. Ancilotto}
\affiliation{\mbox{Dipartimento di Fisica e Astronomia 'Galileo Galilei' and CNISM, Universit\`{a} di Padova, 35131 Padova, Italy}}
\affiliation{CNR-IOM Democritos, 265–34136 Trieste, Italy}
\author{M. Modugno}
\affiliation{\mbox{Depto. de Fis\'ica Te\'orica e Hist. de la Ciencia, Universidad del Pais Vasco UPV/EHU, 48080 Bilbao, Spain}}
\affiliation{IKERBASQUE, Basque Foundation for Science, 48013 Bilbao, Spain}
\author{F. Minardi}
\affiliation{Istituto Nazionale di Ottica, CNR-INO, 50019 Sesto Fiorentino, Italy}
\affiliation{\mbox{LENS and Dipartimento di Fisica e Astronomia, Universit\`{a} di Firenze, 50019 Sesto Fiorentino, Italy}}
\affiliation{Dipartimento di Fisica e Astronomia, Universit\`{a} di Bologna, 40127 Bologna, Italy}
\author{C. Fort}
\affiliation{Istituto Nazionale di Ottica, CNR-INO, 50019 Sesto Fiorentino, Italy}
\affiliation{\mbox{LENS and Dipartimento di Fisica e Astronomia, Universit\`{a} di Firenze, 50019 Sesto Fiorentino, Italy}}
\begin{abstract}
We report on the production of a $^{41}$K-$^{87}$Rb dual-species Bose--Einstein condensate with tunable interspecies interaction and we study the mixture in the attractive regime, i.e. for negative values of the interspecies scattering length $a_{12}$. The binary condensate is prepared in the ground state and confined in a pure optical trap. We exploit Feshbach resonances for tuning the value of $a_{12}$. After compensating the  gravitational sag between the two species with a magnetic field gradient, we drive the mixture into the attractive regime. We let the system to evolve both in free space and in an optical waveguide. In both geometries, for strong attractive interactions, we observe the formation of self-bound states, recognizable as quantum droplets. Our findings prove that robust, long-lived droplet states can be realized in attractive two-species mixtures, despite the two atomic components possibly experiencing different potentials.
\end{abstract}
\maketitle
\section{Introduction}
\label{sec:Introduction}
Ultracold atomic mixtures, formed by atoms of the same species in different spin states, or as different isotopes or elements, enable us to address a wide variety of problems in many-body physics. They have been largely used to investigate phase-transitions \cite{Catani2008,Danshita2014,MikiOta2019,Penna_2019}, multi-component superfluidity \cite{Inguscio2002,Salomon2014, Roy2017}, topological defects \cite{Cornell2004,Hoefer2011,Yao2016}, magnetism \cite{altman_phase_2003,kuklov_counterflow_2003,Ueda2013}, ultracold chemistry \cite{Jin2012} and impurity and polaron physics \cite{Catanimpurity,hu_bose_2016,jorgensen_observation_2016}. {Most of these phenomena are accessible thanks to the ability to control the sign and strength of interactions, opening the way to the study of unconventional matter phases. For instance, in two-component Fermi gases, by tuning the interparticle scattering length close to a Feshbach resonance, it has been possible to explore the Bose--Einstein condensate (BEC)--BCS crossover \cite{varenna}, and in multicomponent Fermi systems, even more exotic scenarios have been discussed which are relevant for high-Tc superconductivity \cite{Iskin206,Tajima_PRB_2019,Yerin2019,Salasnich_prb_2019,Tajima_cmd}. On the other hand, in bosonic mixtures, Feshbach resonances have been recently exploited} to produce and study novel quantum phases, arising by competing interaction forces. Indeed, as pointed out by D. Petrov \cite{Petrov}, the interplay between attractive inter- and repulsive intra-species couplings can give rise 
to the formation of liquid-like ``quantum droplets''.
These were first observed in single-species dipolar gases \cite{kadau2016,FerrierBarbut2016,Schmitt2016,Ferlaino2016,FerrierBarbutJPB2016,Wenzel2017}, wherein the attractive dipole--dipole interaction plays the role of the interspecies interaction, and then in homonuclear mixtures of $^{39}$K \cite{Cheiney2018,Cabrera2018,Semeghini2018} and heteronuclear mixtures of $^{41}$K-$^{87}$Rb \cite{DErrico2019}. In the  latter cases, as originally proposed \cite{Petrov, PetrovAstrakharchik}, the system is characterized only by isotropic contact interactions, 
allowing for a simple theoretical description of the droplet state. Currently, these states, which represent a unique example of ultradilute isotropic liquid, are the subject of intense theoretical and experimental research aimed to unveil their peculiar properties \cite{Kartashov2018,Astrakharchik2018,Cappellaro2018,2Dvortexquantumdroplets_2018,Reimann_PRL_2019,Ferioli2019,Kartashov2019,ferioli2019dynamical,julidaz2020quantum,parisi2020quantum,tylutki2020collective}.
\section{Experiment}
\label{sec:Experiment}
\begin{figure}[htb!]
\centering
\includegraphics[width = .75\textwidth]{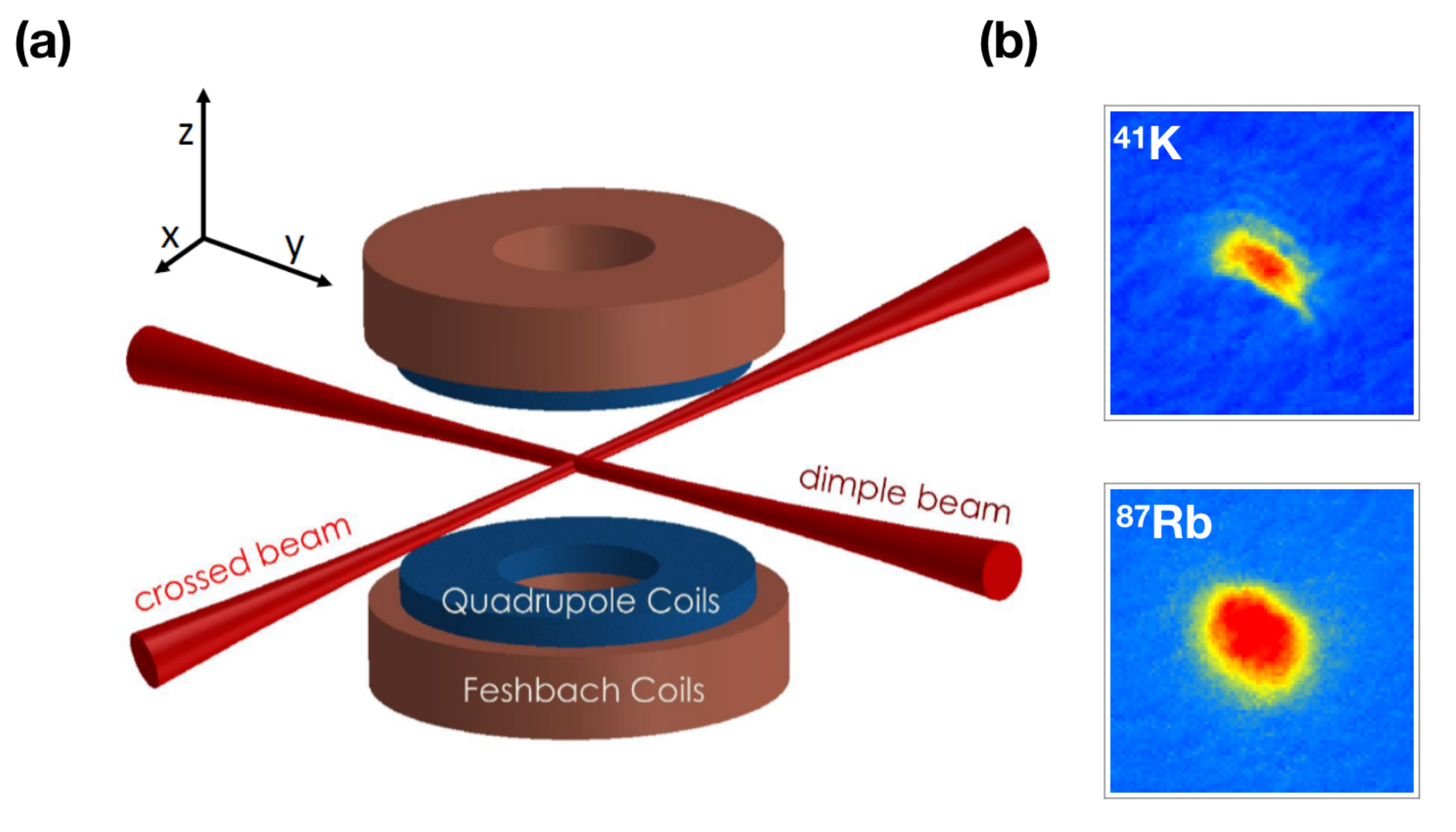}
\caption{\label{fig:experiment} (\textbf{a}) Schematics of the experiment, showing the optical dipole trap (ODT) beams, quadrupole and Feshbach coils. (\textbf{b}) Absorption images of the dual-species Bose--Einstein condensate (BEC).  Images are taken after 18~ms of TOF for $^{41}$K and 20.5~ms for $^{87}$Rb. $N_\mathrm{K} \sim 2 \times 10^4$ and  $N_\mathrm{Rb} \sim 6 \times 10^4$. During TOF expansion, the Fesbach field is set at $77.5$~G, corresponding to $a_{12} \simeq 255 a_0$.}
\end{figure}

The experimental set-up for the production of  $^{41}$K-$^{87}$Rb  BEC has been described in ~\cite{Burchianti2018}. Briefly, after a first cooling stage in a two-species magneto-optical trap, we optically pump both $^{41}$K and $^{87}$Rb in the low-field-seeking state $\left|F=2,m_{F}=2\right\rangle$, and then we load them in an hybrid potential, consisting of a magnetic quadrupole plus an optical dipole trap (ODT) (see  Figure~\ref{fig:experiment}a). The quadrupole field is produced by the same set of coils we use for the magneto-optical trap, while the ODT is made by two far-off-resonance laser beams at a wavelength of 1064~nm. One beam (labeled as dimple beam in Figure~\ref{fig:experiment}a) comes along the $\hat{y}$ axis, while the other (labeled as crossed beam in Figure~\ref{fig:experiment}a) forms an angle of 67.5$^\circ$ with the dimple beam in the $xy$ plane, and is inclined at an angle of 16$^\circ$ with respect to the same plane. The dimple and crossed beams, delivered by the same laser with a maximum total power of 3~W, are focused at the center of the quadrupole with averaged waists of 90 and 70~$\mu$m, respectively.\\
\label{sec:Experiment}

In the hybrid potential $^{87}$Rb is first magnetically and then optically evaporated, while $^{41}$K is almost sympathetically cooled by elastic collisions with $^{87}$Rb. Following the procedure described  in ~\cite{Burchianti2018}, we end the cooling sequence with a degenerate mixture in the pure optical trap, with trap frequencies ($\nu_{x},\nu_y, \nu_z) \simeq (110, 70, 150)$~Hz for $^{41}$K, and $(75, 45, 85)$~Hz for $^{87}$Rb, corresponding to powers of 400 and 160~mW for the dimple and the crossed beam, respectively. Due to the different masses ($m_1\simeq 0.47 m_2$, with $1 \rightarrow$ $^{41}$K and $2 \rightarrow$ $^{87}$Rb) the trapping potential minima
of $^{41}$K and $^{87}$Rb 
are separated
along the vertical $\hat{z}$ axis of about 15~$\mu$m. This implies that the two atomic samples barely touch in the trap at the end of the evaporation. At this point, we drive the mixture into its absolute ground-state, where easily accessible Feshbach resonances allow for tuning the interspecies scattering length $a_{12}$ \cite{Thalhammer2008}. To this end, we transfer first $^{87}$Rb and then $^{41}$K into the $\left|F=1,m_{F}=1\right\rangle$ state by sweeping 6.8~GHz microwave and a 269~MHz radio frequency in the presence of a vertical polarizing magnetic field of 8~G. The transfer efficiency is 75\% for $^{41}$K and 85\% for $^{87}$Rb, while the residual fraction of both species in the $\left|F=2,m_{F}=2\right\rangle$ state is optically removed. We finally produce a dual-species BEC in the ground-state with about $10^{5}$ atoms. We find that, for the efficient sympathetic cooling of $^{41}$K, the number of condensed $^{87}$Rb atoms is three to four times larger than the one of 
$^{41}$K. 
In Figure~\ref{fig:experiment}b we show absorption images of the dual-species BEC after time-of-flight (TOF) expansion. The images of $^{41}$K and $^{87}$Rb are recorded using the same optical system and temporally spaced by $2.5$~ms. The intraspecies scattering lengths of $^{41}$K and $^{87}$Rb are both positive ($a_{11}=65 a_0$ \cite{DErrico2007} and $a_{22}=100.4 a_0$ \cite{Marte2002}), while their interspecies scattering length $a_{12}$ is magnetically controlled by an homogeneous magnetic field $B$ produced along the $\hat{z}$ axis by an auxiliary set of coils (labeled as Feshbach coils in Figure~\ref{fig:experiment}a). Spanning the value of $B$ around $70$~G, $a_{12}$ varies from negative to positive values, allowing us to explore the whole phase diagram of the quantum mixture, as discussed in Section~\ref{sec:Tuning the interspecies interaction}. 
\section{Tuning the Interspecies Interaction}
\label{sec:Tuning the interspecies interaction}
\begin{figure}[htb!]
\centering
\includegraphics[width = .75\textwidth]{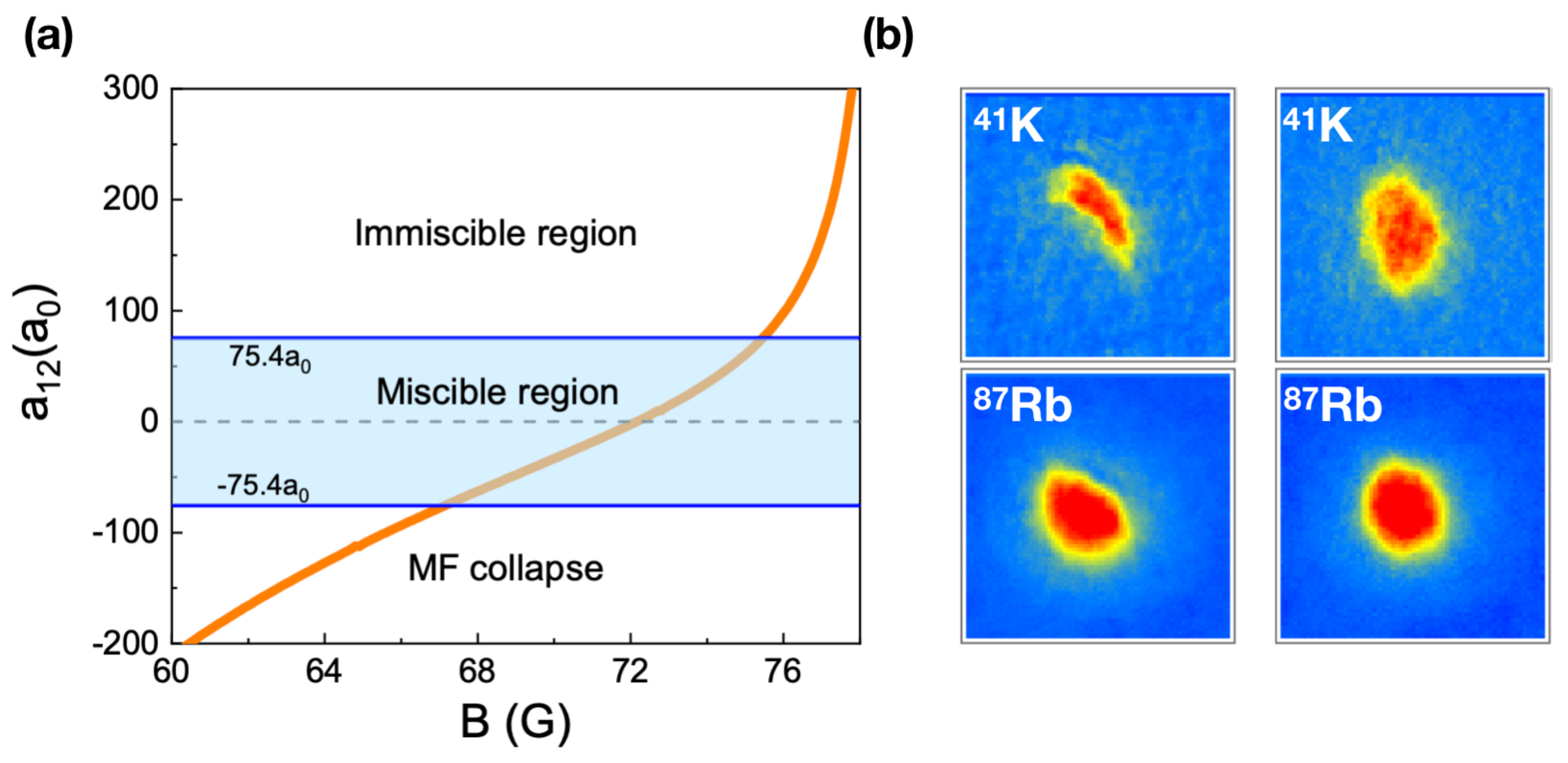}
\caption{ \label{fig:a12vsB} (\textbf{a}) Calculated interspecies scattering length $a_{12}$  for $^{41}$K and $^{87}$Rb in the $\left|F=1,m_{F}=1\right\rangle$ state as a function of the magnetic field $B$. (\textbf{b}) Absorption images of the dual-species BEC. First column: immiscible regime. During time-of-flight (TOF) expansion, the Feshbach field is set at $77.5$~G, corresponding to $a_{12} \simeq 255 a_0$. Second column: miscible regime. During TOF expansion, the Feshbach field is set at $73$~G, corresponding to $a_{12} \simeq 10 a_0$.}
\end{figure}

 Within the Thomas--Fermi approximation, an interacting bosonic mixture is completely described in terms of the coupling constants: 
 $g_{ij}={2\pi \hbar^{2} a_{ij}}/{m_{ij}}$, with  
$m_{ij}={m_{i}m_{j}}/{\left(m_{i}+m_{j}\right)}$ 
 ($i,j=1,2$) \cite{Riboli2002}. For a homogeneous system, the ground-state can be 
 a uniform mixture of the two components if $|{g_{12}}| < \sqrt{g_{11}g_{22}}$ or phase-separated if $g_{12} > \sqrt{g_{11}g_{22}}$. For strong enough attractive interspecies interactions, i.e. $g_{12} <- \sqrt{g_{11}g_{22}}$,  consistent with the mean-field (MF) theory, the system is unstable and the mixture is expected to collapse (for our mixture the critical point corresponds to $a^{\mathrm{c}}_{12}=-75.4a_0$). However, quantum fluctuations, arising by the zero-point motion of the Bogoliubov excitations, stabilize the system against the MF collapse and may give rise to the formation of liquid-like quantum droplets \cite{Petrov}, as discussed in Section~\ref{sec:Attractive regime}. 
  In the experiment, we exploit Feshbach resonances for tuning the value of $a_{12}$.  In Figure~\ref{fig:a12vsB}a we show the calculated value of $a_{12}$ \cite{Simoni2008,Thalhammer2009} as a function of $B$ in the range from $60$ to $78$~G, in between two Feshbach resonances. Moving from the left to the right side, we explore the MF unstable region, where droplets may form; the miscible region close to the zero-crossing at $72$~G, where $a_{12}$ vanishes; and finally, for enough repulsive interactions, the immiscible region. As pointed in Section~\ref{sec:Experiment}, the two species hardly overlap in the ODT. Despite this, the effect of the interactions can be unveiled in TOF. Indeed, after switching off the trap, the $^{41}$K and $^{87}$Rb clouds start to expand and to spatially overlap. In Figure~\ref{fig:a12vsB}b we show absorption images of the dual-species BEC expanding in the immiscible (first column) and miscible regimes (second column). In the first case the Feshbach field is set at $77.5$~G corresponding to $a_{12} \simeq 255 a_0$. Due to the strong repulsive interspecies interaction, the two clouds repel each other and the density distribution of each species is affected by the presence of the other. In the second case, the Feshbach field is set at $73$~G, corresponding to $a_{12} \simeq 10 a_0$. The two clouds weakly interact during the expansion, and their density distributions are almost the same as the ones observed for single-species BECs.
\section{Compensating for the Gravitational Sag}
\label{sec:Magnetic compesation}
\begin{figure}[htb!]
\centering
\includegraphics[width = .75\textwidth]{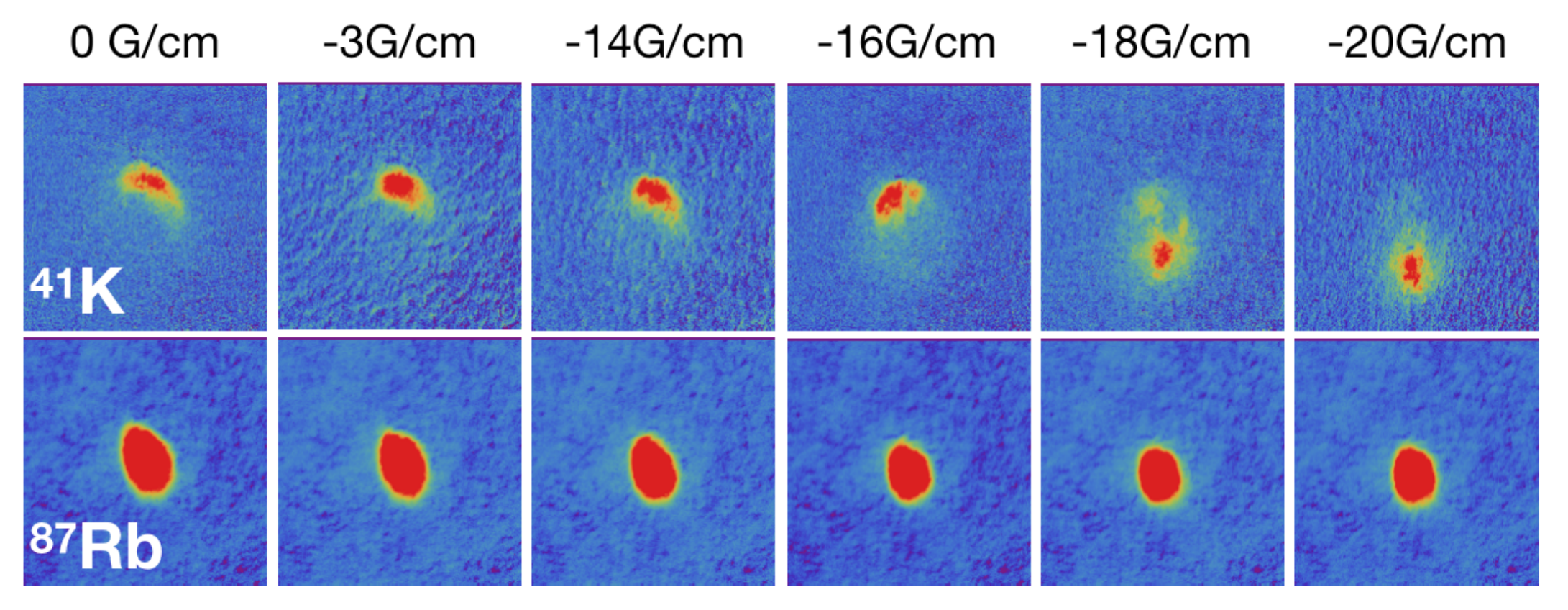}
\caption{ \label{fig:compensation} Absorption images of the dual-species BEC for different values of $b_z$. During TOF expansion, the Feshbach field is set at $77.5$~G corresponding to $a_{12} \simeq 255 a_0$. 
}
\end{figure} 
Studying the role of the interactions when the two species are optically confined requires one to compensate for the gravitational sag. To that end, we apply a vertical magnetic field gradient $b_z$, generated by the quadrupole coils, and we exploit the different magnetic moments of $^{41}$K and $^{87}$Rb in the region of magnetic field explored for tuning $a_{12}$. For 60~G~$\leq B\leq$~78~G the magnetic moment $z$-components of both species are almost constant:
$\mu_1 = 0.83 \;\mu_B$ and $\mu_2= 0.52\;\mu_{B}$, with $\mu_{B}$ the Bohr magneton. Thus, the potential felt by the $i$-th species along the $\hat{z}$ axis is given by
\begin{equation}
\label{eq1}
U_{z,i}=
\frac12 m_i\omega_{i,z}^2 z^2+m_i g z - \mu_i b_z z,
\end{equation}
where $\omega_{i,z}$ is the angular trap frequency of the $i$-th species due to the pure optical potential ($\omega_{i,z} \propto \sqrt{\alpha_i I_0/m_i}$, with $\alpha_i$ the atomic polarizability of the $i$-th species and $I_0$ the maximum trap laser intensity). The minimum $z_{0,i}$ of $U_{z,i}$ can be easily calculated
\begin{equation}
\label{eq2}
z_{0,i}=\frac{ \mu_i b_z - m_i g} {m_i \omega_{i,z}^2},
\end{equation}
and by imposing the condition $\Delta z=z_{0,1}-z_{0,2}$=0, it follows:
\begin{equation}
\label{eq3}
b_z=g\frac{m_1 \alpha_2-m_2 \alpha_1}{\mu_1 \alpha_2-\mu_2 \alpha_1}.
\end{equation}
We find that in our mixture a magnetic field gradient $b_z$ of about -17 G/cm fulfills Equation~\ref{eq3} ($ \alpha_1 / \alpha_2 \simeq 0.88$). The effect of this gradient is to move down the minimum of the trapping potential, $z_{0,i}$, in the direction of the gravity. 
The magnetic moment of the lighter $^{41}$K is larger than that of $^{87}$Rb, as a result its displacement due to $b_z$ is larger. This allows it to compensate for the differential gravitational sag and to overlap the two species.
In Figure~\ref{fig:compensation}, we show the $^{41}$K and $^{87}$Rb  density distributions, after TOF expansion, for different values of $b_z$. After preparing the quantum mixture in the ground-state, we first ramp the magnetic field $B$ up to $72$~G, we increase the power of the dimple beam to 1~W, and then we apply the  magnetic gradient. The recompression of the trap has a twofold effect: on one hand, it reduces the starting value of $\Delta z$, and on the other, it compensates for the lowering of the trap depth due to $b_z$. Before releasing the mixture from the trap, $B$ is set at $77.5$~G corresponding to $a_{12} \simeq 255 a_0$; thus, the two clouds are phase-separated during TOF. For $b_z=0$,  the $^{41}$K cloud is above the $^{87}$Rb one because $z_{0,1} > z_{0,2}$ in the trap. By increasing the value of $b_z$ we observe that the positions of both clouds change, and after overtaking the expected value to compensate for the gravitational sag, their positions are reversed, namely the $^{41}$K cloud is below the $^{87}$Rb one.
\section{The Attractive Regime}
\label{sec:Attractive regime}
 In this section we study the quantum mixture in the attractive regime for weak and strong interspecies interactions, respectively, $a^{\mathrm{c}}_{12}<a_{12}<0$ and $a_{12}<a^{\mathrm{c}}_{12}$. After overlapping the two species, we decrease the magnetic field $B$ below $72$~G with a linear ramp of 30~ms. We let the system evolve either in free-space or in a horizontal waveguide. In the latter case, we remove the confinement due the crossed beam by linearly decreasing its power to zero in 10~ms. 
Within the MF theory, the binary gas is stable in the weakly attractive regime, while it is expected to become unstable beyond the critical value $a^{\mathrm{c}}_{12}$. However, this simplified picture fails once we take into account the effect of quantum fluctuations, as detailed in the following.
\subsection{Phase Diagram: Free Space}
The MF energy density of a homogeneous mixture of bosons reads:
\begin{equation}
\label{eq4}
{\cal E} _{\rm MF}=\frac1 2 g_{11} n_1^2+\frac1 2 g_{22} n_2^2+g_{12} n_1 n_2,
\end{equation}
with $n_i$ being the atomic density of the $i$-th species (we assume $g_{11},g_{22}>0$), which can be also expressed in terms of  $\delta g=g_{12}+\sqrt{g_{11} g_{22}}$:
\begin{equation}
\label{eq5}
{\cal E} _{\rm MF}=\frac12(n_1 \sqrt{g_{11}}- n_2 \sqrt{g_{22}}) \,^2+n_1 n_2 \delta g.
\end{equation}
We consider the case $\delta g <0$ with $| \delta g| \ll {g_{ii}}$ ($i=1,2$). The system stability requires that the first term in Equation~(\ref{eq5}) to be zero, fixing the equilibrium ratio between the densities of the two species $n_1 / n_2 = \sqrt{g_{22} / g_{11}}$
(for our mixture $n_1 / n_2 \simeq 0.85$). Furthermore,  the second term in Equation~(\ref{eq5}) also has to be minimized. This implies that the atomic densities increase, leading the system to collapse, unless the MF attractive energy term is balanced by an additional energy contribution.
For a binary mixture, the first-order correction to the MF energy, the so called Lee--Huang--Yang (LHY) term \cite{LeeHuangYang}, is approximately
\begin{align}
{\cal E} _{\rm LHY} \simeq \frac{8}{15 \pi^2} \left(\frac{m_1}{\hbar^2}\right)^{3/2}
\!\!\!\!\!\!(g_{11}n_1)^{5/2}
\left(1+z^{3/5} x\right)^{5/2},
\label{eq6}
\end{align}
with $z=m_2/m_1$ and $x=g_{22} n_2/( \,g_{11} n_1) \,$. This equation extends the analytic formula given in ~\cite{Petrov}, valid for homonuclear mixtures ($z=1$) to the heteronuclear case ($z \ne 1$) \cite{Minardi2019}. The MF and LHY energy contributions have opposite signs, i.e. ${\cal E} _{\rm MF} \propto \delta g < 0$ and ${\cal E} _{\rm LHY} > 0$, the latter depending on the repulsive interspecies coupling constants $g_{ii}$. Further, since ${\cal E} _{\rm MF} \propto n^{2}$ and ${\cal E} _{\rm LHY} \propto n^{5/2}$ there exists a finite value of the atomic density minimizing the energy ($n \approx n_1 \approx n_2$). Thus, the competition between the LHY and MF contributions gives rise to the formation of self-bound droplets. The droplet equilibrium density can be calculated by setting  $P=-{\cal E}+\sum_{i}({\partial {\cal E} }/{\partial n_i})\,n_i=0$, with $P$  the pressure and $\cal E = {\cal E} _{\rm MF}+{\cal E} _{\rm LHY}$  the total density energy. This condition gives:
\begin{align}
n^{0}_{1}&=\frac{25\pi}{1024\, a_{11}^3}\frac{\delta g^2}{g_{11}g_{22}}
\left(1+z^{3/5} x\right)^{-5}
\nonumber\\
n^{0}_{2}&=n^{0}_{1}\sqrt{g_{11} / g_{22}};
\label{eq7}
\end{align}
therefore, for approximately equal intraspecies scattering lengths it follows that $n^{0} \propto (\delta g/g)^2 a^{-3}$.

Until now, we have neglected any finite-size and surface effects. However in a liquid droplet, the density is not uniform but it reaches its maximum at the center and drops to zero at the edges. Such a density gradient has a cost in terms of kinetic energy, which may have dramatic effects in small droplets. Indeed, due to the surface tension, the total energy may shift from negative to positive values, driving again the transition to the gas phase. Thus, for any fixed value of $\delta a = a_{12}-a^{\mathrm{c}}_{12}$, there exists a critical atom number $N_{\mathrm{c}}$ below which the droplet is unstable and evaporates \cite{Petrov}.

\begin{figure}[htb!]
\centering
\includegraphics[width = .60\textwidth]{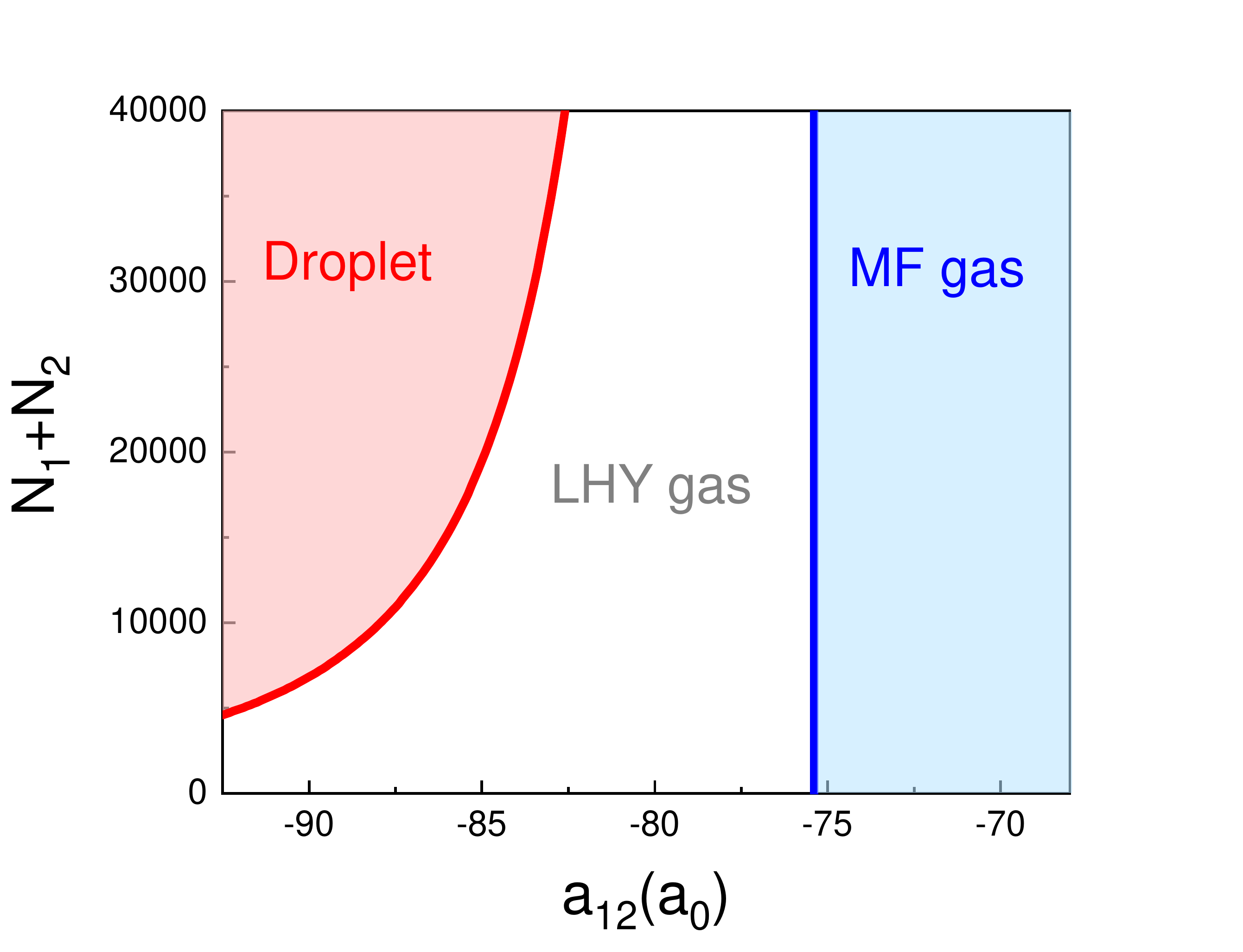}
\caption{\label{fig:phasediagram} Phase diagram of a $^{41}$K-$^{87}$Rb attractive mixture, in free space, as a function of $N_1+N_2$ and $a_{12}$. For $ \delta g >0$  ($a_{12}>a^{\mathrm{c}}_{12}$) the gas is stable within the MF theory. For $ \delta g <0\, (a_{12}<a^{\mathrm{c}}_{12})$, the MF energy is balanced by the Lee--Huang--Yang (LHY) term, and the system forms a LHY gas. For a sufficiently large atom number and sufficiently strong interactions, a self-bound droplet forms.}
\end{figure}

 In Figure~\ref{fig:phasediagram}, we show the phase diagram of a $^{41}$K-$^{87}$Rb attractive mixture in free space as a function of the interspecies scattering length $a_{12}$ and the total atom number $N_1$+$N_2$, with $N_1$ ($N_2$ ) being the atom number of $^{41}$K ($^{87}$Rb) and $N_1 / N_2=0.85$.
\begin{figure}[htb!]
\centering
\includegraphics[width = .70\textwidth]{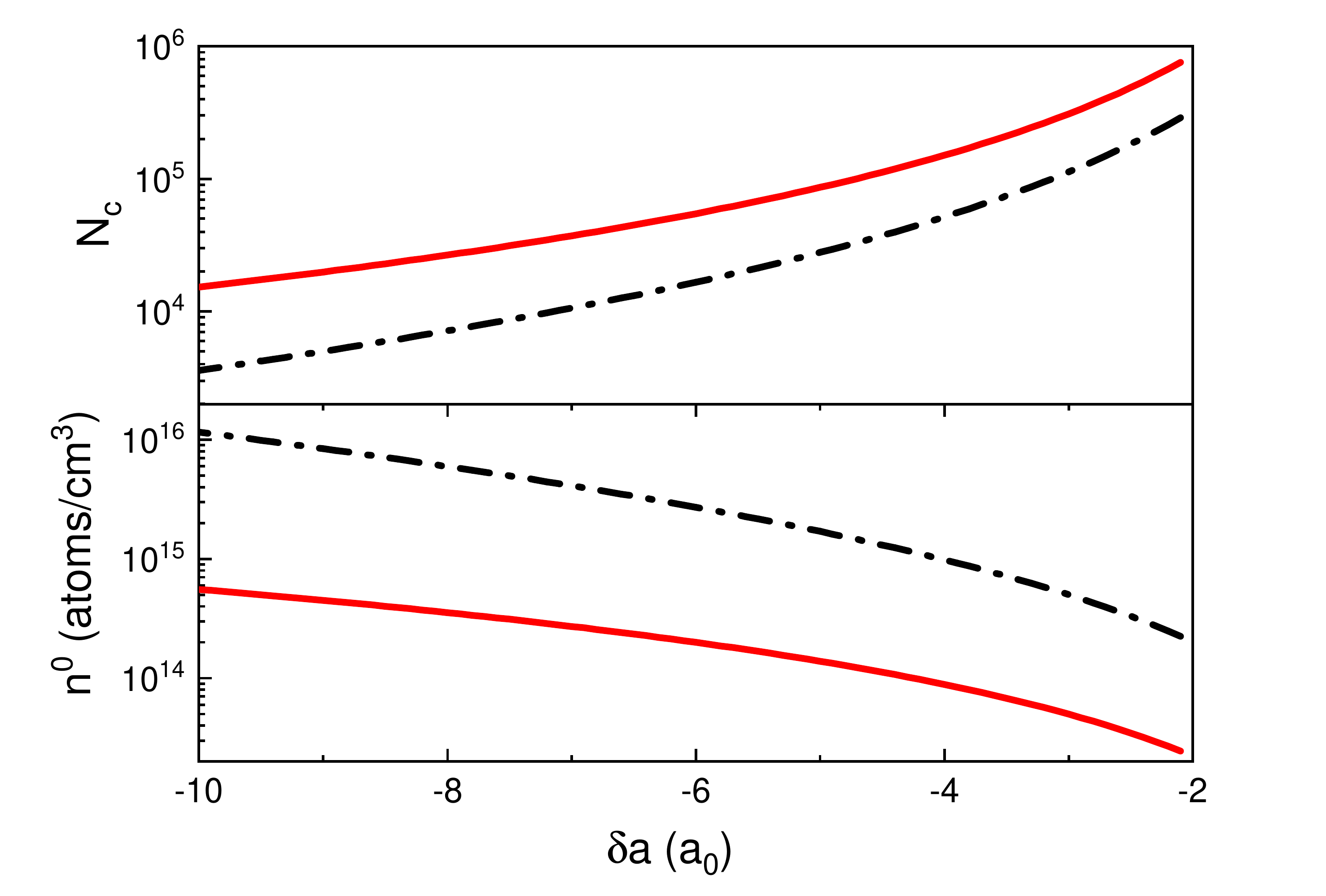}
\caption{\label{fig:dropletdensity} Calculated critical atom number $N_{\mathrm{c}}$ (top panel) and droplet density (bottom panel) in logarithmic scale as a function of $ \delta a$ for the $^{41}$K-$^{87}$Rb mixture (solid red lines) and the $^{39}$K spin mixture (dash-dotted black lines).}
\end{figure} 

For $ \delta g >0$ ($a_{12}>a^{\mathrm{c}}_{12}$), the gas is stable at the MF level, while for $ \delta g <0$ ($a_{12}<a^{\mathrm{c}}_{12}$), beyond the MF threshold for collapse, the mixture is stabilized by quantum fluctuations. If the number of particles is below $N_{\mathrm{c}}$, the system is a so-called LHY gas, while above $N_{\mathrm{c}}$ a droplet forms.  By increasing the strength of the attractive interaction, the droplet binding energy increases, counteracting the surface tension, as a result, $N_{\mathrm{c}}$ decreases. 
In Figure~\ref{fig:dropletdensity} we show, for a $^{41}$K-$^{87}$Rb mixture in free space, the calculated values of the droplet density $n^{0}$ and of the critical atom number $N_{\mathrm{c}}$ as a function of $ \delta a$. For comparison, we also plot the corresponding quantities for a spin mixture of $^{39}$K (in this case $a_{11}$ is magnetically tuned, while $a_{22}$ and $a_{12}$ are constant \cite{Cheiney2018,Cabrera2018,Semeghini2018, Ferioli2019}). We observe that the density of $^{41}$K-$^{87}$Rb droplets is at least one order of magnitude smaller than the density of the $^{39}$K ones, while the value of $N_{\mathrm{c}}$ is comparable for small $ \delta a$. A lower density, which in the case of $^{41}$K-$^{87}$Rb can be ascribed to the larger values of the intraspecies scattering lengths $a_{ii}$ (see Equation~(\ref{eq5})),  can substantially increase the droplet lifetime $\tau_{\mathrm{life}}$. In strongly interacting mixtures the main sources of atomic losses are inelastic three-body collisions, thus $\tau_{\mathrm{life}} \propto n^{-2}$. As a result, a 10-fold density reduction corresponds to a 100-fold longer lifetime.

\subsection{Observation of $^{41}$K-$^{87}$Rb Droplets in Free Space}
 We study the evolution of the attractive $^{41}$K-$^{87}$Rb mixture in free space, after removing the trapping potential. Once the trap is turned off, the atoms drop under the effect of gravity and may exit  the region where the Feshbach field is homogeneous. In principle, a droplet formed by $N_1+N_2$ atoms could be magnetically levitated by a vertical magnetic field gradient $b^{\mathrm{lev}}_z$, satisfying the condition: 
 \begin{equation}
\label{eq8}
b^{\mathrm{lev}}_z=g\frac{N_1 m_1+N_2 m_2}{N_1 \mu_1 + N_2 \mu_2}.
\end{equation}
Thus, a levitation gradient $b^{\mathrm{lev}}_z$ of about 18 G/cm, i.e. in between the levitation gradients of the two species, would hold the droplet against gravity. However, we find that for $b^{\mathrm{lev}}_z >5$~G/cm,  the  bound state breaks up, due to the different magnetic forces acting on the two components. Therefore, we apply a levitation gradient of only 1~G/cm, which, although too small to compensate for gravity, slightly reduces the variation of $B$ experienced by the droplet during the fall. Furthermore, it allows us to spatially discriminate between bound and unbound components, as discussed below.
 
\begin{figure}[htb!]
\centering
\includegraphics[width = 0.85\textwidth]{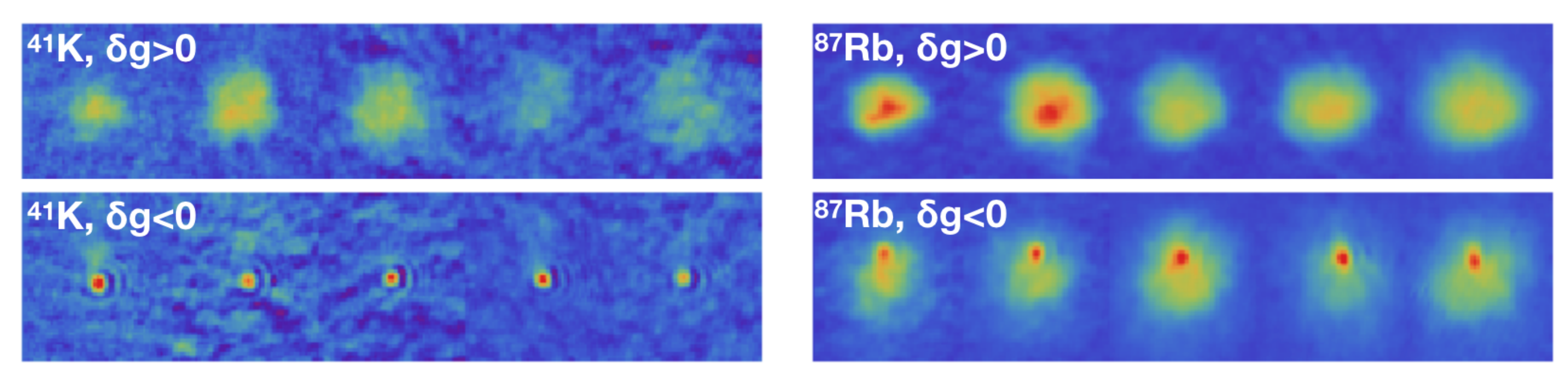}
\caption{\label{fig:dropletfreespace} (\textbf{a}) Absorption images (frame size: 190 x 190 $\mu$m) of $^{41}$K and $^{87}$Rb for different values of TOF. For $^{41}$K from left to right:
TOF=$17.5$, $19.5$, $21.5$, $23.5$, $25.5$~ms. For $^{87}$Rb from left to right: TOF=$20$, $22$, $24$, $26$, $28$~ms. Top panels correspond to $a_\mathrm{12} \simeq -18  a_0$ ($\delta g >0$), and bottom panels to $a_\mathrm{12} \simeq  -85  a_0$ ($\delta g <0$).}
\end{figure}

In Figure~\ref{fig:dropletfreespace} we show absorption images of $^{41}$K and $^{87}$Rb  for different values of TOF. The top panels correspond to the weakly attractive regime ($a_\mathrm{12} \simeq -18  a_0$) and the bottom ones to the strongly attractive regime ($a_\mathrm{12} \simeq  -85  a_0$). In the former case the sizes of both $^{41}$K and $^{87}$Rb clouds increase with TOF, as expected for a gas. In the latter case, instead, a fraction of both species bounds, forming a small and dense sample whose size, in the case of $^{41}$K, does not exceed the imaging resolution ($5\;\mu$m) on a timescale of several tens of ms, indicating the formation of a droplet state. The different size and shape of the bound component in $^{87}$Rb is due the imaging procedure: since the image of $^{87}$Rb is acquired $2.5$~ms after the $^{41}$K one, the bound state has been already dissociated at the time of $^{87}$Rb imaging. This implies that direct information on the formed droplet can be extracted only by the density distribution of $^{41}$K. For both species the fraction of unbound atoms appears as an halo surrounding the bound state (bottom panels 
in Figure~\ref{fig:dropletfreespace}).
\\The presence of this halo is more evident for $^{87}$Rb, since we generally have an excess of unbound $^{87}$Rb (we recall that the $^{87}$Rb atom number is three to four times larger than the $^{41}$K one, while in the droplet $N_1 / N_2=0.85$). An unbound fraction in both species can be also due to the presence of residual thermal components.  The position of the halo is shifted with respect to the droplet: in particular, the 
$^{41}$K halo is upshifted while the $^{87}$Rb one is downshifted, due to the different magnetic moments of the two unbound components.
The experimental results shown in Figure~\ref{fig:dropletfreespace} are consistent with the phase diagram in Figure~\ref{fig:phasediagram} since we estimate to have $N_1+N_2 \simeq 3.3 \times 10^4$ atoms in the bound state \footnote{we evaluate an uncertainty of 30\%  in the estimation of $N_i$  for both atomic species, due to the imaging calibration}. This ensures the droplet stability in the strongly attractive regime for $a_\mathrm{12} <  -82  a_0$.

In the experiment, the droplet state persists for 28~ms; after that the sample starts to expand, indicating the transition to the gas phase. We  ascribe this effect mainly to the variation of $a_\mathrm{12}$ during the fall (we estimate a variation in $B$ of $0.17$~G after a TOF of 25~ms, resulting in a change of $2.9 a_0$ in $a_\mathrm{12}$). Our 28 ms observation time is about three times longer than the measured lifetime of $^{39}$K droplets in free space \cite{Semeghini2018}. This difference is caused by the fact that the $^{41}$K-$^{87}$Rb droplets are more dilute than the $^{39}$K ones. Close to the MF threshold for collapse, we have measured a three-body loss coefficient $K_3$ of the order of $10^{-40}$~m$^6$/s \cite{DErrico2019}, whose value is mainly determined by K-Rb-Rb collisions \cite{wacker2016}. This yields, for small values of $\delta a$, an expected  droplet lifetime $\tau_{\mathrm{life}} = K_3^{-1} n^{-2}$ of the order of 1~s.

\subsection{Observation of $^{41}$K-$^{87}$Rb Droplets in a Waveguide}
\begin{figure}[htb!]
\centering
\includegraphics[width = .75\textwidth]{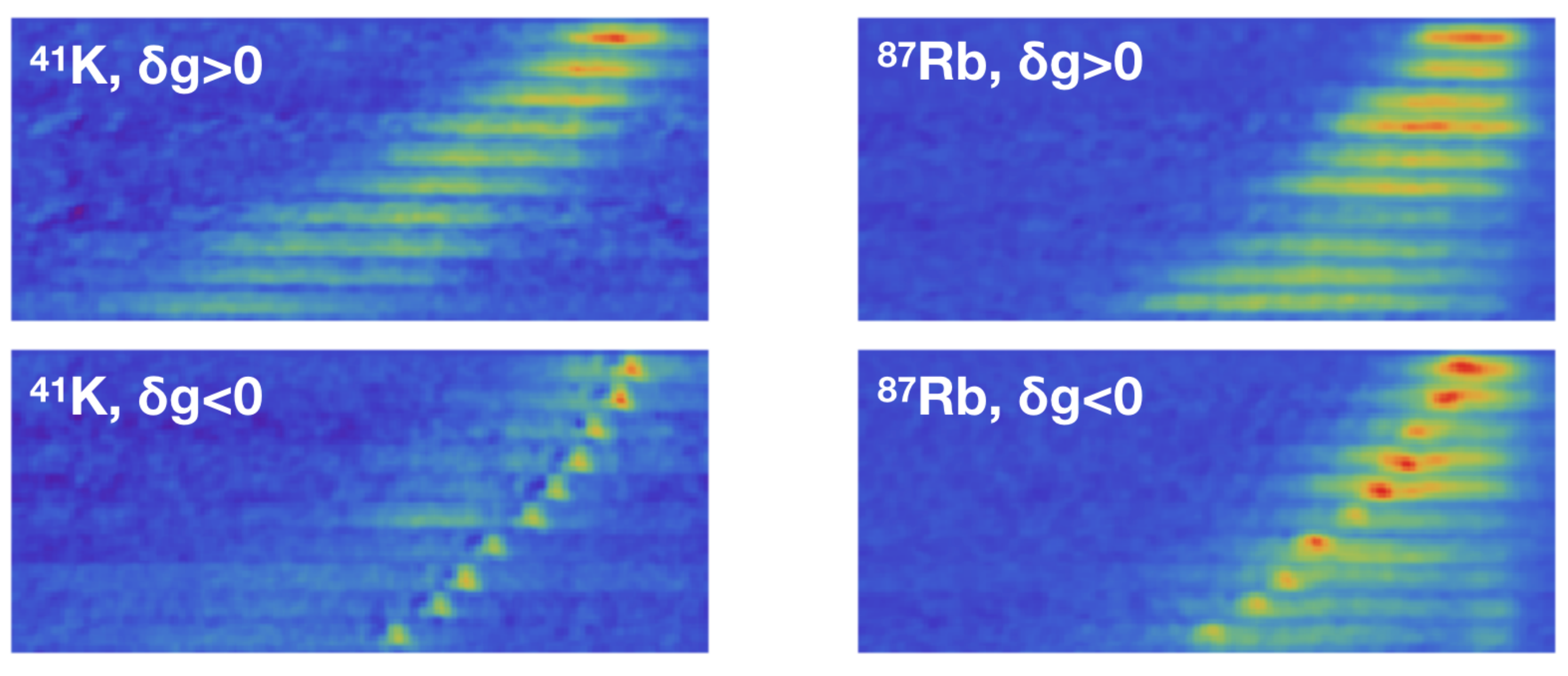}
\caption{\label{fig:dropletwaveguide}  Absorption images (frame size: $34  \times 760$ $\mu$m) ~
of $^{41}$K and $^{87}$Rb for different evolution times $t$ in the waveguide. For $^{41}$K, from top to bottom:
$t$=$15$, $20$, $25$, $30$, $35$, $40$, $45$, $50$, $55$, $60$~ms. For $^{87}$Rb;  from top to bottom:
$t$=$17.5$, $22.5$, $27.5$, $32.5$, $37.5$, $42.5$, $47.5$, $52.5$, $57.5$, $62.5$~ms. Top panels correspond to $a_\mathrm{12} \simeq -12  a_0\, (\delta g >0)$, and bottom panels to $a_\mathrm{12} \simeq  -83  a_0$ ($\delta g <0$).}
\end{figure}

In this section, after preparing the attractive mixture in the ODT, we study its dynamical evolution in the waveguide obtained by removing the crossed beam. In this configuration, the atoms experience an anti-trapping potential with an axial frequency $\nu_{y} \simeq -1.8$~Hz ($-0.6$~Hz) and a radial trapping potential with average frequencies $\nu_{r} \simeq 160$~Hz ($120$~Hz) for $^{41}$K ($^{87}$Rb). The anti-trapping force is due to the magnetic potential given by the Feshbach field and the overlapping magnetic field gradient. Thus, both atomic species move along the guide ($\hat{y}$ direction) under the combined effect of the magnetic force and a residual component of the gravity \footnote{The zero of the quadrupole field, in the horizontal plane, is slightly off-center from the ODT center, in addition the waveguide is vertically tilted of approximately $0.1^{\circ}$--$0.2^{\circ}$. As a result, the atoms under the action of the antitrapping potential start to move along the guide}.
 In Figure~\ref{fig:dropletwaveguide} we show absorption images of $^{41}$K  and $^{87}$Rb for different values of the evolution time $t$, after removing the crossed beam. In  the weakly attractive regime, $a_\mathrm{12} \simeq -11  a_0$ (top panels in Figure~\ref{fig:dropletwaveguide}), we obverse that both atomic components expand along $\hat{y}$, while the positions of their centers of mass move under the effects of both the magnetic force and gravity. As expected, the shift of the center of mass is larger for the $^{41}$K component than for the $^{87}$Rb one, because of its larger magnetic moment. In the strongly attractive regime, $a_\mathrm{12} \simeq  -82.5  a_0$ (bottom panels in Figure~\ref{fig:dropletwaveguide}), in addition to the unbound $^{41}$K and $^{87}$Rb clouds, which move independently each from the other, there exists a fraction of both species bound together by their mutual attraction. The width of this bound state does not increase in time within our imaging resolution. Further, its center-of-mass follows a trajectory in between the ones of the unbound single-species components, and can be easily distinguished from those for sufficiently long evolution times. From the center-of-mass motions of the bound and the unbound components we have extracted the ratio $N_1/N_2$ in the bound state, confirming the result predicted by the theory \cite{DErrico2019}. 
 \\Taking into account our experimental atom number, stable droplets form for $a_\mathrm{12}<-82 a_0$. However, the phase diagram in the waveguide is more complex than in free space. Due to the combined effect of the LHY term and the dispersion along the guide, stable "solitonic" solutions may also exist for $-82 a_0<a_\mathrm{12}<a^{\mathrm{c}}_{12}$ \cite{Cappellaro2018}. Despite this, we do not observe bound states in the aforementioned range of interactions. We ascribe this effect to the procedure used in the experiment to form the droplet: first the initial size of the mixture is adjusted to match the droplet size (smaller than the one of the solitonic ground state); then the system dynamics is induced in a non adiabatic way by removing the crossed beam.
 
\begin{figure}[htb!]
\centering
\includegraphics[width = 0.75\textwidth]{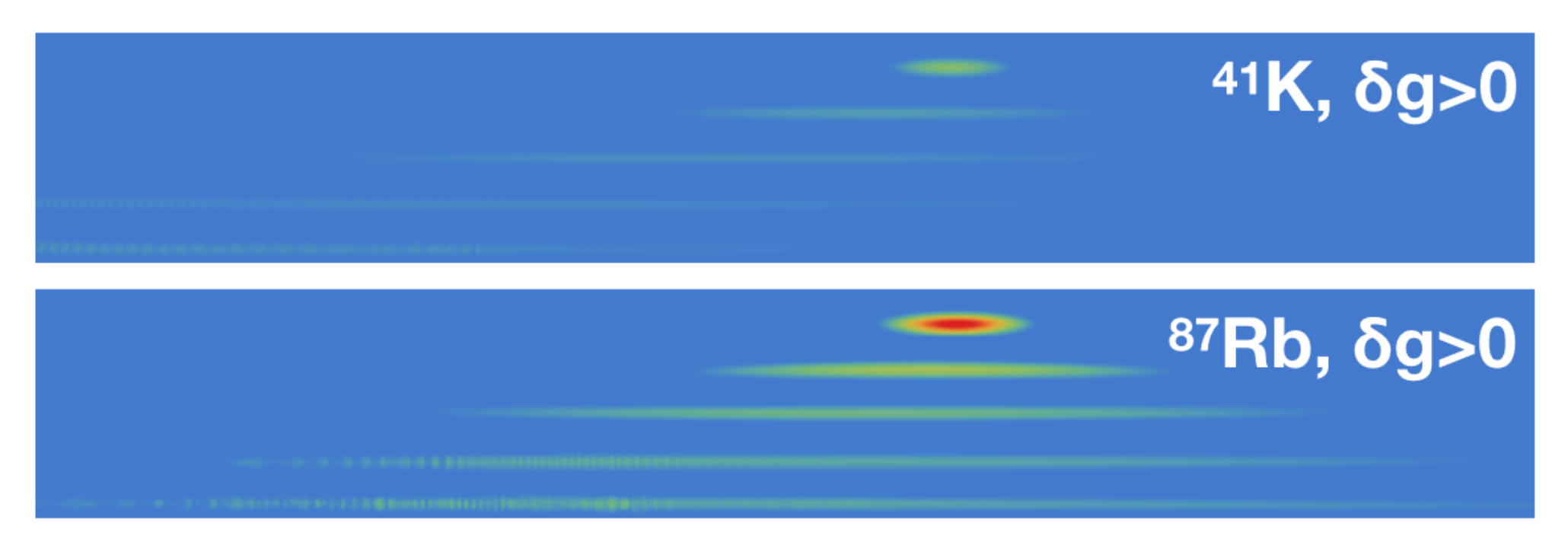}
\caption{\label{fig:simulazione1} Simulated density distributions (frame size: 6  x 155 $\mu$m) of $^{41}$K and $^{87}$Rb for $a_\mathrm{12} = -12  a_0$ ($\delta g >0$),
at different evolution times $t$ in the waveguide. For both species, from top to bottom  $t= 0,10,20,30,40$~ms; $N_1 = 1.5 \times 10^3$ and $N_2 = 4 \times N_1$.}
\end{figure}

\begin{figure}[htb!]
\centering
\includegraphics[width = 0.75\textwidth]{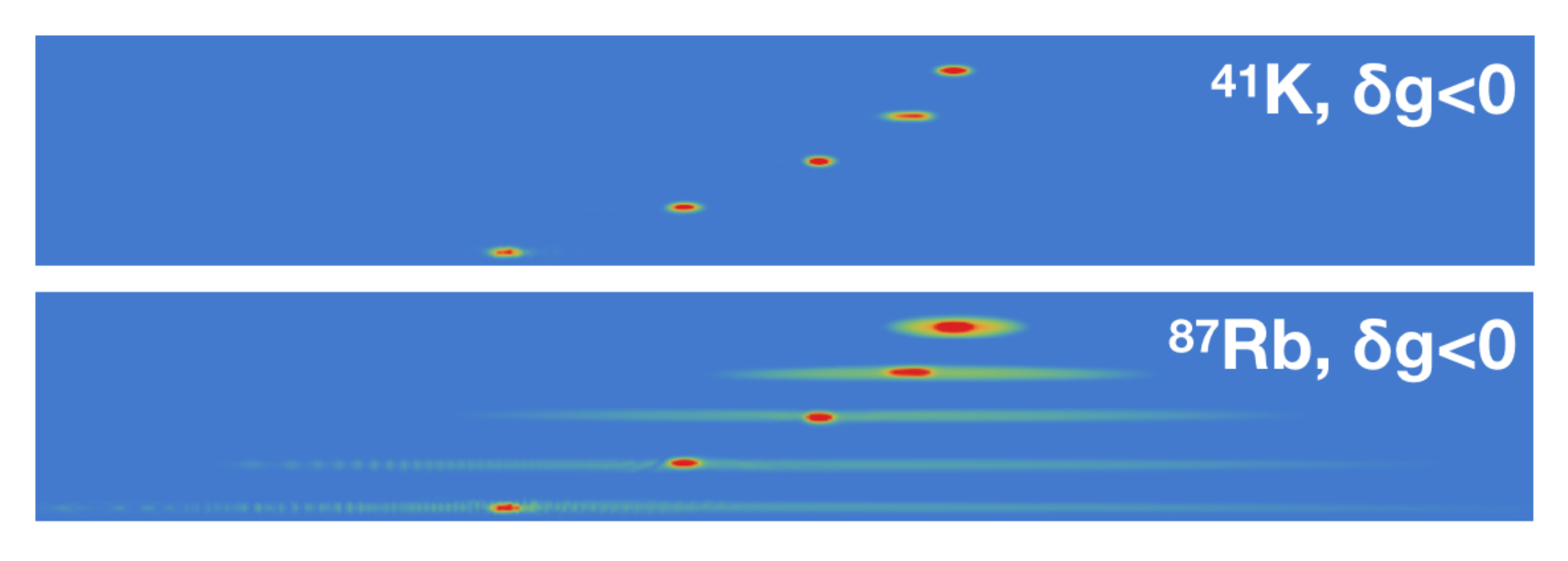}
\caption{\label{fig:simulazione2} Simulated density distributions (frame size: 6 x 155 $\mu$m) of $^{41}$K and $^{87}$Rb for $a_\mathrm{12} = -83  a_0$ ($\delta g <0$),
at different evolution times $t$ in the waveguide.  For both species, from top to bottom  $t= 0,10,20,30,40$~ms; $N_1 = 1.5 \times 10^3$ and $N_2 = 4 \times N_1$.}
\end{figure}

  In order to gain more insights into the droplet formation process and its evolution, we simulate the system dynamics by numerically solving two coupled, generalized Gross--Pitaevskii (GP) equations at $T=0$ \cite{Gross,Pitaevskii}, including the LHY correction for hetereronuclear mixtures \cite{Ancilotto2018, Minardi2019}. We use a complete description of the potential experienced by the two species. The simulation is analogous to the experiment: the ground state is prepared in the ODT at a given interspecies interaction and its evolution is triggered by switching off the crossed beam potential. The center-of-mass motion of the atomic samples is obtained by adding to the magnetic potential a bias magnetic field along $\hat{y}$, reproducing in this way the shift of the quadrupole minimum (here we assume the waveguide as perfectly horizontal, neglecting the tilt present in the experiment). To  match the experimental conditions, we consider the case of an unbalanced mixture with  $N_1 = 1.5 \times 10^3$ and $N_2 = 4 \times N_1$.
 In Figure~\ref{fig:simulazione1} and Figure~\ref{fig:simulazione2} we show the simulated density distributions of $^{41}$K and $^{87}$Rb,  as a function of the evolution time $t$ for the weakly ($a_\mathrm{12} = -12  a_0$ ) and strongly attractive regimes ($a_\mathrm{12} \simeq  -83  a_0$), respectively. In the first case, both atomic samples expand, while in the second case a droplet forms. This is composed by the whole minority component, i.e. $^{41}$K, and a fraction of majority component, i.e. $^{87}$Rb, satisfying the condition $N_1 / N_2=0.85$ (the excess of $^{87}$Rb remains unbound). Furthermore, the droplet's trajectory is between the ones of the two unbound components, qualitatively reproducing the experimental results. The simulations also take into account the three-body losses; however, their effect is negligible for the range of parameters considered here. We point out that for stronger attractive interspecies interactions the atomic losses due to $K_3$ may be important. Nevertheless, the droplet, once its atomic population drops below $N_\mathrm{c}$, does not evaporate, but decays into a soliton, which is a stable solution.

\section{Discussion}
\label{sec:Discussion}
We have shown that robust and largely tunable binary condensates can be realized by exploiting two-species mixtures of $^{41}$K and $^{87}$Rb. Despite their different masses, the two species can be spatially overlapped and their mutual interaction can be magnetically tuned by means of Feshabach resonances, allowing one to access the whole phase diagram. Here, we have focused on the attractive regime, where long-lived, self-bound droplets form both in free space and under radial confinement. In view of studying the droplet properties, the  lifetime $\tau_{\mathrm{life}}$ has to be compared with the characteristic time scale of droplet dynamics $\tau$ \cite{Petrov}. Since the ratio ${\tau_{\mathrm{life}}} / {\tau}$ scales as $n^{-1}$, with our mixture, we  expect an effective improvement by a factor 10 with respect to the $^{39}$K case ($n^{0}_{^{41}K^{87}Rb}\sim 10^{-1} n^{0}_{^{39}K}$). Currently, our droplet observation time  in free space is limited by the variation of $a_\mathrm{12}$ during TOF. In the future, we plan to increase this time by slowing down the droplet's fall, thanks to a species-dependent optical potential \cite{Catanimpurity}. The experimental realization of long-lived droplets opens up the possibility of studying their peculiar properties, such as their excitation spectra \cite{Petrov,Cappellaro2018} and the predicted self-evaporation \cite{Petrov,ferioli2019dynamical}.

Finally, our results prove that droplet formation is not hindered by the different trapping potentials experienced by the two species. As confirmed by the droplet dynamics along the guide, the presence of species-dependent forces may be compensated by the attractive interaction. We plan to extend our analysis by measuring the critical value of the differential acceleration between the two species above which the droplet breaks up. This should provide an accurate estimation of its binding energy.

\begin{acknowledgments}
We thank M. Fattori for valuable discussions, M. Inguscio for support and A. Simoni for collision calculations.This research was funded by the European Commission through FET Flagship on Quantum Technologies---Qombs Project (grant number 820419) and by Fondazione Cassa di Risparmio Firenze through project "SUPERACI-Superfluid Atomic Circuits." M.M. acknowledges support from the Spanish Ministry of Science, Innovation and Universities and the European Regional Development Fund FEDER through grant number PGC2018-101355-B-I00 (MCIU/AEI/FEDER, UE), and the Basque Government through grant number IT986-16.
\end{acknowledgments} 

\bibliography{Biblio}
\end{document}